%% file: 0-main.tex
\documentclass{egpubl}
\usepackage{eurovis2024}
\usepackage{booktabs}
\usepackage{xcolor}
\EuroVisEducation              % uncomment for Education contribution
\usepackage[T1]{fontenc}
\usepackage{dfadobe}  

\usepackage{cite}  % comment out for biblatex with backend=biber
\BibtexOrBiblatex
\electronicVersion
\PrintedOrElectronic
\ifpdf \usepackage[pdftex]{graphicx} \pdfcompresslevel=9
\else \usepackage[dvips]{graphicx} \fi

\usepackage{egweblnk}
\title[ Designing Born Accessible Visualization Courses: Challenges and Opportunities of Developing a Remote Curriculum]%
      {Designing Born-Accessible Courses in Data Science and Visualization: Challenges and Opportunities of a Remote Curriculum Taught by Blind Instructors to Blind Students}

\author[J. Seo \& S. O'Modhrain]
{\parbox{\textwidth}{\centering 
        JooYoung Seo$^{1}$\orcid{0000-0002-4064-6012},
        Sile O'Modhrain$^{2}$\orcid{0000-0003-3804-5469},
        Yilin Xia$^{1}$\orcid{0000-0003-2500-0245},
        Sanchita Kamath$^{1}$\orcid{0000-0001-6469-0360},
        Bongshin Lee$^{3}$\orcid{0000-0002-4217-627X},
        and James M. Coughlan$^{4}$\orcid{0000-0003-2775-4083}
        }
        \\
{\parbox{\textwidth}{\centering 
         $^1$University of Illinois at Urbana-Champaign, School of Information Sciences, USA\\
         $^2$University of Michigan, School of Information, USA\\
         $^3$Yonsei University, Republic of Korea\\
         $^4$Smith-Kettlewell Eye Research Institute, USA
       }
}
}

\begin{document}
\input{1-abstract}
\input{2-introduction}
\input{3-related_work}
\input{4-learning_design}
\input{5-results}
\input{6-discussion}
\input{7-conclusion}
\input{8-acknowledgement}

%-------------------------------------------------------------------------
% bibtex
\bibliographystyle{eg-alpha-doi}
\bibliography{references,eurovis_revision}

% biblatex with biber
% \printbibliography                

%-------------------------------------------------------------------------
%Color tables are no longer required for purely electronic publications.
% \newpage
% 
% 
% \begin{figure*}[tbp]
%   \centering
%   \mbox{} \hfill
%   % the following command controls the width of the embedded PS file
%   % (relative to the width of the current column)
%   \includegraphics[width=.3\linewidth]{sampleFig}
%   % replacing the above command with the one below will explicitly set
%   % the bounding box of the PS figure to the rectangle (xl,yl),(xh,yh).
%   % It will also prevent LaTeX from reading the PS file to determine
%   % the bounding box (i.e., it will speed up the compilation process)
%   % \includegraphics[width=.3\linewidth, bb=39 696 126 756]{sampleFig}
%   \hfill
%   \includegraphics[width=.3\linewidth]{sampleFig}
%   \hfill \mbox{}
%   \caption{\label{fig:ex3}%
%            For publications with color tables (i.e., publications not offering
%            color throughout the paper) please \textbf{observe}: 
%            for the printed version -- and ONLY for the printed
%            version -- color figures have to be placed in the last page.
%            \newline
%            For the electronic version, which will be converted to PDF before
%            making it available electronically, the color images should be
%            embedded within the document. Optionally, other multimedia
%            material may be attached to the electronic version. }
% \end{figure*}

\end{document}

\maketitle
\begin{abstract}
   While recent years have seen a growing interest in accessible visualization tools and techniques for blind people, little attention is paid to the learning opportunities and teaching strategies of data science and visualization tailored for blind individuals. Whereas the former focuses on the accessibility issues of data visualization tools, the latter is concerned with the learnability of concepts and skills for data science and visualization. In this paper, we present novel approaches to teaching data science and visualization to blind students in an online setting. Taught by blind instructors, nine blind learners having a wide range of professional backgrounds participated in a two-week summer course. We describe the course design, teaching strategies, and learning outcomes. We also discuss the challenges and opportunities of teaching data science and visualization to blind students. Our work contributes to the growing body of knowledge on accessible data science and visualization education, and provides insights into the design of online courses for blind students.
   %-------------------------------------------------------------------------
   %  ACM CCS 1998
   %  (see https://www.acm.org/publications/computing-classification-system/1998)
   % \begin{classification} % according to https://www.acm.org/publications/computing-classification-system/1998
   % \CCScat{Computer Graphics}{I.3.3}{Picture/Image Generation}{Line and curve generation}
   % \end{classification}
   %-------------------------------------------------------------------------
   %  ACM CCS 2012
   % (see https://www.acm.org/publications/class-2012)
   %The tool at \url{http://dl.acm.org/ccs.cfm} can be used to generate
   % CCS codes.
   %Example:

\ccsdesc[500]{Applied computing~Education}

\printccsdesc
\end{abstract}

\section{Research Questions}

Our research questions aim to understand the effects of teaching data visualization techniques to blind individuals. We will use the pre- and post-surveys to address these key questions:

1. How does teaching data visualization techniques to blind individuals impact their confidence in data visualization?

2. Does a course on data visualization techniques improve the sense of agency in learning for blind individuals?

3. What is the experience of blind individuals with data visualization before the course? 

4. How does prior experience with specific visualizations (such as barplot, heatmap, boxplot, and scatterplot) affect blind individuals' learning and confidence in using those visualizations?

5. How do blind individuals envision using data visualization techniques for their career development and professional growth?

6. Did the course on data visualization techniques meet the expectations of blind individuals in terms of their career development and professional growth?

7. What were the most challenging and rewarding aspects of the course for blind individuals in terms of their learning and professional growth?

8. How has the course on data visualization techniques impacted the career development and professional growth of blind individuals?

These research questions will guide our analysis and interpretation of the survey data to gain insights into the effects of teaching data visualization techniques to blind individuals.

\section{Pre-Survey}
\subsection{Introduction}

This survey aims to gauge your initial self-efficacy, sense of agency, and prior experiences with data science, data visualization, and specific visualizations. We also aim to understand more about your visual impairment, demographic information, screen reader products, refreshable braille display model, education level, learning goals, and career aspirations. Your responses will help us tailor the course to meet your needs. All information collected is confidential and used for educational and research purposes only.

\subsubsection{Demographics}

1. What is your age?: \rule{1cm}{0.15mm}
\\
\\
2. Gender: \\
   - [ ] Male\\
   - [ ] Female\\
   - [ ] Prefer not to say\\
   - [ ] Other: \rule{1cm}{0.15mm}
\\
\\
3. Occupation: \rule{1cm}{0.15mm}\\
   - Field: \rule{1cm}{0.15mm}
\\
\\
4. Education Level: \\
   - [ ] High School or equivalent\\
   - [ ] Associate's Degree\\
   - [ ] Bachelor's Degree\\
   - [ ] Master's Degree\\
   - [ ] Doctorate Degree\\
   - [ ] Other: \rule{1cm}{0.15mm}
\\
\\
5. Major: \rule{1cm}{0.15mm}
\\
\\
6. Minor (if applicable):\rule{1cm}{0.15mm}

\subsubsection{Visual Impairment}

7. How long have you been blind? \rule{1cm}{0.15mm}
\\
\\
8. Can you describe your level of visual impairment in your left eye? \rule{1cm}{0.15mm}
\\
\\
9. Can you describe your level of visual impairment in your right eye? \rule{1cm}{0.15mm}

\subsubsection{Assistive Technologies}

10. What screen reader product(s) do you use?\rule{1cm}{0.15mm}
\\
\\
11. How long have you been using this screen reader? \rule{1cm}{0.15mm}\\
\\
12. What model of refreshable braille display do you use? \rule{1cm}{0.15mm}\\
\\
13. How long have you been using this refreshable braille display?\rule{1cm}{0.15mm}\\

\subsubsection{Prior Experience}

14. Have you had any experience with data science? \\
   - [ ] Yes \\
   - [ ] No \\
\\
15. If yes, please briefly describe your experience with data science. \rule{1cm}{0.15mm} 
\\
\\
16. Have you had any experience with data visualization? \\
   - [ ] Yes \\
   - [ ] No 
\\
\\
17. If yes, please briefly describe your experience with data visualization. \rule{1cm}{0.15mm} \\
\\
18. How have you been accessing data visualization? \rule{1cm}{0.15mm}\\
\\
19. Please score your level of confidence in creating the following types of visualizations on a scale of 1 to 5, with 1 being the lowest confidence and 5 being the highest confidence:\\
   - Barplot: \rule{1cm}{0.15mm} \\
   - Heatmap: \rule{1cm}{0.15mm} \\
   - Boxplot: \rule{1cm}{0.15mm} \\
   - Scatterplot: \rule{1cm}{0.15mm} \\
\newline
20. Please rate your level of confidence for interpreting each visualization on a scale of 1 to 5, with 1 being the lowest confidence and 5 being the highest confidence: \\
   - Barplot: \rule{1cm}{0.15mm} \\
   - Heatmap: \rule{1cm}{0.15mm} \\
   - Boxplot: \rule{1cm}{0.15mm} \\
   - Scatterplot: \rule{1cm}{0.15mm} \\
\\
21. Have you used tactile representations (e.g., embossed or 3D graphs) of the following types of visualizations before? \\
   - [ ] Barplot\\
   - [ ] Heatmap\\
   - [ ] Boxplot\\
   - [ ] Scatterplot\\
\\
22. For each of the following types of visualizations, how well do you believe you know what they look like? Please rate your level of knowledge (1- No knowledge at all, 5 - Completely knowledgeable): \\
   - Barplot: \rule{1cm}{0.15mm} \\
   - Heatmap: \rule{1cm}{0.15mm} \\
   - Boxplot: \rule{1cm}{0.15mm} \\
   - Scatterplot: \rule{1cm}{0.15mm} \\
\\
23. Please rate your current level of computer programming skills (1- None, 5 - Very proficient): \rule{1cm}{0.15mm}\\
\\
24. How many years of programming experience do you have? \rule{1cm}{0.15mm}\\
\\
25. Have you used any of the following data science environments before? For each environment you've used, please specify how long you've been using it.\\
   - [ ] R (Length of usage: \rule{1cm}{0.15mm})\\
   - [ ] Python (Length of usage: \rule{1cm}{0.15mm})\\
   - [ ] SPSS (Length of usage: \rule{1cm}{0.15mm})\\
   - [ ] SAS (Length of usage: \rule{1cm}{0.15mm})\\
   - [ ] MatLab (Length of usage: \rule{1cm}{0.15mm})\\
   - [ ] Stata (Length of usage: \rule{1cm}{0.15mm})\\
   - [ ] MiniTab (Length of usage: \rule{1cm}{0.15mm})\\
   - [ ] Visual Studio Code (Length of usage: \rule{1cm}{0.15mm})\\
   - [ ] Other (Specify: \rule{1cm}{0.15mm}, Length of usage: \rule{1cm}{0.15mm})\\
\\
26. What do you want to learn in this course? Please describe your learning goals. \rule{1cm}{0.15mm} \\
\\
27. How do you envision using data visualization techniques for your career development or professional growth? \rule{1cm}{0.15mm} 
\\

\subsubsection{Preferred Modality for Learning}

28. Which modality or combination of modalities do you prefer for learning data visualization techniques? Please select all that apply: \\
   - [ ] Braille \\
   - [ ] Sonification \\
   - [ ] Text-to-speech \\

\section{Post-Survey}
\subsection{Introduction}
This survey aims to understand your self-efficacy, sense of agency, and experiences with data visualization after the course. We appreciate your honest feedback. Your responses will help us improve future sessions.

\subsubsection{Self-efficacy and Sense of Agency}
1. How much control do you feel you have over your learning process now? (1- No control at all, 5 - Complete control)\\
1 [ ] 2 [ ] 3 [ ] 4 [ ] 5 [ ]

\subsubsection{Course Feedback}
2. What was the most challenging aspect of the course for you?  \rule{1cm}{0.15mm} \\
\\
3. What was the most rewarding part of the course for you?  \rule{1cm}{0.15mm} \\
\\
4. On a scale of 1 to 5, how well did the course meet your expectations in terms of your career development or professional growth? \rule{1cm}{0.15mm}\\
\\
5. Please briefly explain your rating for the last question  \rule{1cm}{0.15mm} \\
\\
6. How has this course impacted your career development or professional growth?  \rule{1cm}{0.15mm}

\subsubsection{Visualization Confidence}
7. How confident are you in your ability to create the following data visualization after this course? (1- Not at all confident, 5 - Very confident)\\
   - Barplot:  \rule{1cm}{0.15mm} \\
   - Heatmap:  \rule{1cm}{0.15mm} \\
   - Boxplot:  \rule{1cm}{0.15mm} \\
   - Scatterplot:  \rule{1cm}{0.15mm} \\
   \\
8. Please briefly explain your rating.  \rule{1cm}{0.15mm}\\
\\ 
9. Please rate your level of confidence in interpreting each of the following visualizations now, after completing the course (1- Not at all confident, 5 - Very confident):
   - Barplot:  \rule{1cm}{0.15mm} \\
   - Heatmap:  \rule{1cm}{0.15mm} \\
   - Boxplot:  \rule{1cm}{0.15mm} \\
   - Scatterplot:  \rule{1cm}{0.15mm} \\

%% file: 1-abstract.tex
\maketitle
\begin{abstract}
  While recent years have seen a growing interest in accessible visualization tools and techniques for blind people, little attention is paid to the learning opportunities and teaching strategies of data science and visualization tailored for blind individuals. Whereas the former focuses on the accessibility and usability issues of data visualization tools, the latter is concerned with the learnability of concepts and skills for data science and visualization. In this paper, we present novel approaches to teaching data science and visualization to blind students in an online setting. Taught by blind instructors, nine blind learners having a wide range of professional backgrounds participated in a two-week summer course. We describe the course design, teaching strategies, and learning outcomes. We also discuss the challenges and opportunities of teaching data science and visualization to blind students. Our work contributes to the growing body of knowledge on accessible data science and visualization education, and provides insights into the design of online courses for blind students.
   %-------------------------------------------------------------------------
   %  ACM CCS 1998
   %  (see https://www.acm.org/publications/computing-classification-system/1998)
   % \begin{classification} % according to https://www.acm.org/publications/computing-classification-system/1998
   % \CCScat{Computer Graphics}{I.3.3}{Picture/Image Generation}{Line and curve generation}
   % \end{classification}
   %-------------------------------------------------------------------------
   %  ACM CCS 2012
   % (see https://www.acm.org/publications/class-2012)
   %The tool at \url{http://dl.acm.org/ccs.cfm} can be used to generate
   % CCS codes.
   %Example:
\begin{CCSXML}
<ccs2012>
<concept>
<concept_id>10010405.10010489</concept_id>
<concept_desc>Applied computing~Education</concept_desc>
<concept_significance>500</concept_significance>
</concept>
</ccs2012>
\end{CCSXML}

\ccsdesc[500]{Applied computing~Education}

\printccsdesc
\end{abstract}

%% file: 2-Introduction.tex
\section{Introduction}
\label{sec:introduction}

Over the past decade, there has been a growing body of research on making data visualization accessible to a broader range of people \cite{kimAccessibleVisualizationDesign2021,leeReachingBroaderAudiences2020,marriottInclusiveDataVisualization2021,seoTeachingVisualAccessibility2023,lee_et_al:DagRep.13.6.81}. The data visualization and accessibility communities, in particular, have actively investigated non-visual data representation and interaction methods for blind and low-vision (BLV) individuals who face extra challenges in directly accessing data visualizations. Techniques that have been explored include data sonification using spatial sound \cite{summersConvertingGraphicalDatavisualizations2019,hoqueAccessibleDataRepresentation2023}, natural language data descriptions \cite{lundgardAccessibleVisualizationNatural2022,choiVisualizingNonVisual2019}, tactile graphics with embossed braille or haptic feedback \cite{brownVizTouchAutomaticallyGenerated2012a,degreefInterdependentVariablesRemotely2021}, and multimodal data representations that combine various methods to supplant visual elements \cite{thompsonChartReaderAccessible2023,sharifVoxLensMakingOnline2022,seoMAIDRMultimodalAccess2023}.

However, the focus has been predominantly on addressing accessibility and usability issues in data visualization systems and techniques. In addition, as Fitzpatrick and colleagues have highlighted, there exists a tendency to inadvertently categorize blind individuals as mere consumers rather than creators of data visualizations \cite{fitzpatrickProducingAccessibleStatistics2017b}.

In contrast, this paper explores the learnability of data science and visualization by, with, and for blind people. We introduce a novel approach to teaching data science and visualization to blind students through an online course, facilitated by blind instructors. We present a two-week summer course, which took place in August 2023, involving nine blind learners with diverse professional backgrounds from banking, to managers of databases for state and federal administration, and to neuroscience research.

This paper addresses the following research questions: 
\begin{enumerate}
    \item How can we scaffold the learning of data science and visualization for blind students in an online setting?
    \item How much can blind students learn about data science and visualization in a two-week online course?
    \item What are the challenges and opportunities of teaching data science and visualization to blind students, especially by blind instructors?
\end{enumerate}

We detail the course design, teaching strategies, and learning outcomes, and discuss the challenges and opportunities encountered in teaching data science and visualization to blind students. Our work contributes to the expanding knowledge base on accessible data science and visualization education, and offers insights into designing online courses for blind students.

%% file: 3-related_work.tex
\section{Related Work}
\label{sec:related-work}

\subsection{Teaching Statistics and Visualizations to Blind Learners}
\label{sec:teaching-statistics}

Teaching statistics and data visualization to blind students poses significant challenges due to the heavy reliance on visual representations of data and concepts. Previous research has explored various methods and tools to make statistics and data science education more accessible and inclusive for blind learners.

Gibson and Darron \cite{gibsonTeachingStatisticsStudent1999}, for example, conducted an exploratory case study with one blind student to identify effective adaptations for making statistical concepts accessible without visual aids. They provided the student with tactile graphics and braille handouts created using a Perkins Brailler and swell paper. The student achieved above average scores on exams and assignments, suggesting that blind students can learn foundational statistics with appropriate accommodations.

Marson et al. \cite{marsonTeachingIntroductoryStatistics2013} reviewed the literature on teaching statistics to blind students and offered practical advice based on their own experiences. They recommended useful classroom aids such as Nemeth Code for math braille, tactile models of distributions, and kinesthetic explanations. They also discussed assessment strategies and potential pitfalls for instructors, such as being overly sympathetic or lowering expectations.

Godfrey and Loots \cite{godfreyAdviceBlindTeachers2015} provided a unique perspective as blind PhD statisticians and university lecturers reflecting on their own experiences learning and teaching statistics. They emphasized the importance of communication and direct questions to elicit blind students' needs. They also suggested specific accommodations such as accessible texts, recorded lectures, quality embossed images, and accessible software like R or SAS.

Our work contributes to this knowledge base on accessible data science education with the insights into designing online courses for blind students.

\subsection{Accessible Tools for Data Science Education}
\label{sec:accessible-tools}

The most comprehensive work on making statistical software accessible to blind students and professionals has been carried out by Jonathan Godfrey, a senior faculty member at Massey University in New Zealand who is himself blind. Godfrey outlined the accessibility issues faced by blind users when accessing statistical software and explained how R has proven to be the most accessible option \cite{godfreyStatisticalSoftwareBlind2013}. He introduced a new R package called BrailleR \cite{braille2018}
% \footnote{https://github.com/ajrgodfrey/BrailleR} 
aimed at converting graphical output to text descriptions for blind users. He also proposed some minor improvements in R to further enhance its accessibility, such as alternate formats for documentation and functions that provide text representations of graphs. Godfrey has also collaborated with other BLV scholars to extend this work. For example, Fitzpatrick, Godfrey, and Sorge presented a method for producing accessible statistical diagrams in R for blind users \cite{fitzpatrickProducingAccessibleStatistics2017b}. They used the gridSVG \cite{murrell2014gridsvg} package in R
% \footnote{https://cran.r-project.org/web/packages/gridSVG/index.html}
to generate detailed SVG graphics files encoded with semantic information \cite{fitzpatrickProducingAccessibleStatistics2017b}. They then used the DIAGcess JavaScript library \cite{godfrey2018accessible, sorge2015end}
% \footnote{https://npm.io/package/diagcess} 
to enable interactive web-based exploration of the diagrams through screen readers, synchronized highlighting, and magnification. They demonstrated the method for various types of graphs, such as bar charts, histograms, box plots, and time series graphs. 

In addition, Godfrey and his colleagues extended the previous work by proposing a hierarchical navigation model to support exploring the accessible diagrams \cite{godfreyAccessibleInteractionModel2018b}. They provided summary descriptions of the full chart first, then allowed users to drill down into specific chart components. They also discussed the limitations of screen readers and the need for additional modalities such as braille and sonification.

More recently, Seo et al. have developed a multimodal access and interactive data representation (MAIDR) system to enhance the accessibility and inclusivity of data science and visualization education for BLV learners \cite{seoMAIDRMultimodalAccess2023,seoMAIDR2024}. The MAIDR system provides customizable multimodality modes (i.e., braille, text, and sonification) to learners with varying degrees of visual disabilities. It works with four types of graphs (bar plot, heatmap, box plot, and scatter plot).

Taken together, the synthesis of prior research underscores that the application of embossed tactile graphs for scaffolding visual schema concepts, combined with the provision of accessible statistical software and packages (e.g., R, BrailleR). This is accomplished by leveraging a multimodal data representation system such as MAIDR \cite{seoMAIDRMultimodalAccess2023,seoMAIDR2024}, which offers a viable pathway for blind individuals to achieve competencies in statistics, data science, and visualization comparable to those of their sighted counterparts.

%% file: 4-learning_design.tex
\begin{figure*}
  \centering
  \includegraphics[width=\linewidth]{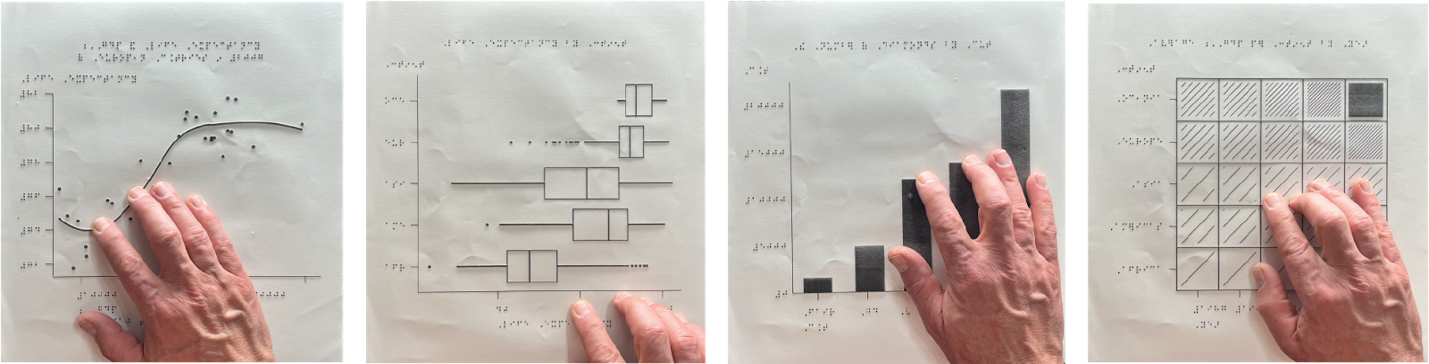}
  \caption{Tactile graphics in swell paper shared with participants: (A) scatter plot, (B) box plot, (C) bar plot, and (D) heatmap.}
  \label{fig:tactile_graph}
\end{figure*}

\section{Learning Design}
\label{sec:learning_design}

The course presented in this paper was specifically designed for blind individuals to explore data science and visualization using R. It was developed as part of an initiative funded through the Rehabilitation and Engineering Research Center (RERC) on Blindness and Low Vision at the Smith-Kettlewell Eye Research Institute.  It included four interactive sessions, each three-hour long, delivered via Zoom and led by a blind professor who’s an expert in using R for data science work. We allocated three hours per session to allow for a balance between lecture, hands-on practice, and Q\&A time. Since both instructors and students used screen readers, we had to be mindful of the time needed to navigate through the code and the screen reader's speech output. While a course of this duration  cannot cover everything in the vast field of data science, our main focus was on data visualization and exploration in non-visual ways. The concepts taught, while centered around R, can be easily applied to other programming languages such as Python.

Our target learners were those who require data visualization skills in their academic or professional journey. This included faculty, undergrad/graduate students, programmers, analysts, engineers, and scientists. Due to the personalized nature of this course, we could only accommodate up to 10 participants.
Selection was based on the application and immediate needs of the participants.
Participation in pre- and post-course interviews and surveys, as well as completing a short assignment after each session, was requested for research purposes. The study was conducted in accordance with the ethical guidelines and regulations set forth by the approved Institutional Review Board (IRB) at both the Smith-Kettlewell Eye Research Institute and the University of Illinois at Urbana-Champaign.

\subsection{Objectives}
\label{sec:objectives}

Our primary objective in designing and teaching this course was to equip blind individuals with the tools and knowledge necessary to independently create and interpret data visualizations, contributing to a more inclusive data literacy movement. By doing so, we aimed to increase participants' effectiveness in their professional careers and thereby to bridge the representation gap in the data science field.  Specifically, the course addressed:

\begin{itemize}
    \item How to set up the data science environment, including installing R and Visual Studio Code (VSCode), and configuring them  accessible to blind learners.
    \item Basic concepts of data science workflows.
    \item How to create and explore multimodal data visualization in an accessible, non-visual fashion using ggplot2 \cite{ggplot2016} and MAIDR packages.
    % \footnote{https://ggplot2.tidyverse.org} 
    % \yilin{shall we say R ggplot2 and JavaScript MAIDR packages?}
    % \jooyoung{re: we used R maidr binder so it was an R package.}
    \item How to effectively share and communicate with sighted people through their own data and visualization.
    % \yilin{ through accessible data visualizations?}
    % \jooyoung{re: not necessarily accessible visualization because they produced them via ggplot2 package.}
\end{itemize}

\subsection{Sessions and Materials}
\label{sec:learning_settings_and_materials}
The course consisted of four sessions across two weeks (Aug 7 and 9 for the first week and Aug 14 and 16 for the second week, with each session lasting three hours (from 8-11am in US Pacific Time), resulting in a total of 12 hours. Besides, office hours will be made available upon request from participants.

We provided the participants with the course materials, including the technical setup guide, lecture slides, and R scripts, in advance via our GitHub website at \url{https://jooyoungseo.github.io/a11y_ds/}. The technical setup was prepared by the primary instructor and the course content slide decks were adapted from the open-source ``Data Science in a Box'' curriculum \cite{TidyverseDatasciencebox2024}. We also provided tactile graphics in swell paper to participants in advance to help them understand the basic concepts of data visualization (Figure~\ref{fig:tactile_graph}), which is available on the course website for download.

\noindent\textit{Session 1: Setting Up The Environment.}
We used VSCode as our primary programming integrated development environment (IDE) because it has been proven one of the most accessible programming code editors for screen reader users \cite{seoCodingNonVisuallyVisual2023}. Participants learned how to set up their environment, including installing R, VSCode, and configuring assistive devices for accessibility.
% \yilin{not sure if we want to put the course website here}
% \jooyoung{re: the course website is under construction so I think we can skip it.}

\noindent\textit{Session 2: Data Science Fundamentals.}
In this session, participants were introduced to the basic concepts of data science, including data importing, tidying, transforming, visualizing, and modeling. This workflow was inspired by Hadley Wickham’s R for Data Science \cite{wickhamDataScienceImport2016}.

\noindent\textit{Session 3: Data Visualization and Exploration.}
Participants explored data visualization and how to interpret data non-visually. In advance of the course, tactile graphics illustrating four common data visualizations (Figure~\ref{fig:tactile_graph}) -- Bar Plots, Scatter Plots, Box Plots and Heatmaps -- were prepared and shipped to participants. Behind the selection of these four Cartesian-based visualizations was the consideration of their widespread use in various fields, their distinct statistical properties (e.g., frequency distribution; correlation and regression; interquartile range and outliers; cross-table and chi-square), and their compatibility with MAIDR system \cite{seoMAIDR2024}. During the course, we guided learners to get familiar with the basic concepts of these four visualizations and then participants learned how to create some of these data visualizations using the ggplot2 package. Later, participants were taught to explore the visualized data using dynamic sonification and a refreshable braille display using MAIDR package \cite{seoMAIDRMultimodalAccess2023,seoMAIDR2024}.

\noindent\textit{Session 4: Data Visualization Do It Yourself.}
In the final session, our goal was to have participants brought their own data and to create their own visualizations. Although this turned out to be an overambitious goal, we were able to guide participants through the process of importing an external dataset and creating a visualization of their choice following the tidyverse principles \cite{wickhamWelcomeTidyverse2019}. Due to the wide range of participants' technological background and required assistance (Section~\ref{sec:variability}), we were not able to cover all the participants' personalized data visualization needs within the time frame of the course. However, participants were encouraged to continue practicing and exploring data visualization on their own after the course.
% \yilin{seems a sudden stop here, should we mention about the short period of the training, difficult to find an way to resolve participants programming bugs i.e., the error is too diverse? }

% we have one-on-one pre- course interviews, which were conducted in the week beginning July 31.

% Course Schedule.
% Interviews and screening: Tues Aug 1 and Wed Aug 2 (2 hours each).
% \begin{itemize}
%     \item Session 1: Monday Aug 7 (3 hours).
%     \item Session 2: Wednesday Aug 9 (3 hours).
%     \item Session 3: Monday Aug 14 (3 hours).
%     \item Session 4: Wednesday Aug 16 (3 hours).
% \end{itemize}

\begin{table*}[!ht]
  \caption{Overview of participant demographics, and refreshable braille displays and screen readers employed in the study. The numerical values in the names of the braille devices typically indicate the number of braille cells they contain. An exception is the HIMS Inc BrailleSense 6, which has 32 cells.}
  \label{tab:participant}
  \scriptsize
  \centering
  \begin{tabular}{p{0.4cm} p{3cm} p{2cm} p{0.4cm} p{1.2cm} p{3.2cm} p{3.2cm}}
    \toprule
    PID & Braille Device Name       & Gender     & Age & Education & Major                              & Screen Reader \\
    \midrule
    P01 & Freedom Scientific Focus 40 & Female    & 30  & Doctorate & Cognitive Psychology               & JAWS \\
    P02 & VarioUltra 40              & Male      & 38  & Bachelor & Computer Animation and Media Arts & VoiceOver \\
    P03 & VarioUltra 20              & Male    & 30  & Master & Political Science                & JAWS \\
    P04 & HIMIS inc BrailleSense 6   & Male      & 28  & Master & Computer Science                 & JAWS, NVDA, VoiceOver \\
    P05 & Humanware Brailliant BI 40X & Cisgender Female & 37 & Doctorate & Special Education             & NVDA \\
    P06 & Freedom Scientific Focus 40 & Male      & 48  & Doctorate & Political Science                 & JAWS \\
    P07 & Humanware Brailliant BI 20X & Cisgender Female & 29 & Master & Psychology              & JAWS \\
    P08 & Freedom Scientific Focus 40 Blue & Cisgender Female & 23 & Bachelor & Flip Performance   & JAWS \\
    P09 & Freedom Scientific Focus 40 & Female  & 50  & Doctorate & Communication                     & JAWS, NVDA \\
    \bottomrule
  \end{tabular}
  %\newline
  %\raggedright{\textit{Note: The numerical values in the names of the braille devices typically indicate the number of braille cells they contain. An exception is the HIMS Inc BrailleSense 6, which has 32 cells.}}
\end{table*}

\subsection{Course Team and Students}
\label{sec:instructors_and_students}
The course team consists of one primary instructor, one assistant instructor, one teaching assistant (TA), and one course supervisor. Two instructors are blind while the TA and course supervisor are sighted. All team members are located in the USA.

JooYoung Seo was the primary Instructor for this course. Dr. Seo is an assistant professor in the School of Information Sciences, University of Illinois at Urbana-Champaign. He is a certified RStudio data science educator, a blind scientist, and has contributed to various open-source data science projects to improve their accessibility. He is also closely collaborating with industry partners, including Posit (formerly RStudio) and Microsoft VSCode team, to enhance the accessibility of data science tools and technologies. 

Sile O'Modhrain was the co-instructor. Dr. O'Modhrain is also a blind scientist. She is an associate professor of music in the School of Music, Theatre \& Dance, and associate professor in the School of Information at the University of Michigan. 

Teaching Assistant Yilin Xia is a doctoral student from the School of Information Sciences at the University of Illinois at Urbana-Champaign, with a focus on accessible data science. He was responsible for hosting office hours and helping resolve technical problem encountered by the students.

Course Supervisor Dr. Coughlan, director of the Smith-Kettlewell Rehabilitation Engineering Research Center (RERC) on Blindness and Low Vision, provided overall guidance in developing this course.

% \jooyoung{It'd be great if we could briefly talk about Yilin's TA role in this course.}

Participants for this course were recruited primarily from the Blind Academics list, a private email forum for BLV professionals and graduate students who actively engaged in research. Participants were required to:

\begin{itemize}
  \item Be 18 years or older.
  \item Be a US citizen or current resident.
  \item Be an everyday screen reader user (e.g., JAWS, NVDA, VoiceOver, ORCA) using speech output and/or braille display.
  \item Have access to a laptop or desktop that can run Zoom, VSCode, and R.
  \item Have access to a refreshable braille display or braille notetaker that can be connected to a computer and screen reader.
  \item Have a basic understanding of computer programming, such as variables, functions, loops, and package and data loading.
  \item Have an immediate need for data visualization skills in their academic or professional careers.
\end{itemize}

The nine participants in the course had an average age of 34.78 years (\textit{SD} = 8.73), with ages spanning from 23 to 50 years old (Table~\ref{tab:participant}). Gender distribution among the group were two females, four males, and three individuals identifying as cisgender females. Their educational backgrounds varied, with qualifications ranging from bachelor's degrees to master's and PhDs, in disciplines including Cognitive Psychology, Political Science, and Computer Science. The participants came from a range of work environments including banking, education administration, and industrial and academic research labs.

In terms of their visual impairments, the majority of participants had been blind since infancy. Two participants experienced visual impairments after turning five years old, and P05 was born with low vision and became entirely blind at 25. %Besides, participant P05 was born with low vision and became fully blind at the age of 25. 
In terms of assistive technology, the average starting age for using a screen reader was 16.33 (\textit{SD} = 6.65) with a duration of use ranging from just 10 to over 30 years. Most have been using braille for about five years, although P03 has only recently begun to use braille displays. To optimize their experience with the course, we gathered information on their preferred screen readers  and operating systems in advance. Seven participants preferred using Job Access With Speech (JAWS) on Windows, two of whom also utilized Non-Visual Desktop Access (NVDA), while P02 exclusively used VoiceOver on a Mac. 

Overall, this course demonstrated the presence of an active community of blind professionals with a declared need to visualize their own data for analysis purposes, but also a need to confidently and independently generate visual representations of their data to communicate with their sighted peers.
%
% \yilin{probably we should delete sentence after ",but also.." the later sounds more like a qualitative result, which can not be easily extracted from the participant info }

\subsection{Assessments}
\label{sec:assessments}

To determine how participants' confidence level in data visualization changed after the course, we conducted a pre- and post- self-report survey and interviews (refer to Appendix). For the four visualizations covered in the course (i.e., bar plot, box plot, heatmap, and scatter plot), pre- and post-assessments of confidence in creating and interpreting them were measured using Likert scales~\cite{bertram2007likert}. Semi-structured interviews were also conducted along with the pre- and post-surveys for deeper interpretation.

%% file: 5-results.tex
\section{Results}
\label{sec:results}

\begin{table*}[!ht]
  \caption{The changes of confidence levels. * indicates a statistically significant difference.}
  \label{tab:quant_results}
  \scriptsize
  \centering
  \begin{tabular}{p{2cm} p{2cm} p{3cm} p{1.2cm} p{1cm} p{3.2cm} p{.2cm}}
    \toprule
    Chart Type & Tasks       & Median Scores     & \textit{p} & Z & IQR \\
    \midrule
    Bar Plot    & Creation  & increased from 2 to 5*    & 0.016 & 0.0 & remained at 1 \\
    Heatmap     & Creation  & increased from 1 to 3     & 0.088 & 6.0 & decreased from 2 to 1 \\
    Box Plot    & Creation  & increased from 1 to 4*    & 0.016 & 0.0 & increased from 1 to 2 \\
    Scatter Plot & Creation & increased from 2 to 4*    & 0.017 & 0.0 & remained at 1 \\
    \midrule
    Bar Plot & Interpretation & remained at 5 & 0.317 & 0.0 & remained at 0 \\
    Heatmap & Interpretation & increased from 2 to 3  & 0.221 & 3.0  & increased from 1 to 2 \\
    Box Plot & Interpretation & increased from 3 to 5* & 0.017 & 0.0 & increased from 0 to 1 \\
    Scatter Plot & Interpretation & remained at 4 & 0.414 & 1.5 & remained at 2 \\
    \bottomrule
  \end{tabular}
\end{table*}

\subsection{Quantitative Analysis}
\label{sec:statistical-analysis}

We used the Wilcoxon Signed-Rank Test \cite{woolson2007wilcoxon} to compare pre- and post-test scores for each metric, suitable due to the paired nature of our data and the non-normal distribution of score changes.

%\noindent\textit{Confidence in Graph Creation and Interpretation.}
\label{sec:confidence-graph-creation-interpretation}
Significant improvements in participants' confidence levels were observed in several areas of graph creation and interpretation as a result of attending this data science course (see Table~\ref{tab:quant_results}). 
These results indicate statistically significant improvements in participants' confidence in creating bar plots, box plots, and scatter plots, and interpreting box plots as a result of attending the data science summer course. While some metrics showed non-significant changes, the overall trend suggests a positive impact of the course on enhancing data science skills among participants. Future work will involve recalculating effect sizes accurately to fully understand the magnitude of these changes.

\subsection{Qualitative Analysis}
\label{sec:qualitative-analysis}

We collected qualitative feedback from participants to understand their experiences and identify areas for improvement. The feedback was analyzed using thematic analysis and the following three common themes and patterns emerged: the challenges of audio-based learning, the difficulties of setting up and using the software environment, and the benefits of the course for enhancing their data literacy and confidence. We present each theme below with illustrative quotes from the participants.

\subsubsection{Challenges of Audio-based Learning}
\label{sec:challenges-audio-based-learning}

Students reported that one of the major challenges of the course was to divide their auditory attention between multiple sources of speech, such as their own screen reader, the instructor's screen reader, and the instructor's spoken presentation. This made it hard for them to follow the code examples and the data visualizations.

\begin{quote}
\textit{You’re juggling the verbal information of the code and like understanding it. Plus the spatial information and the. And then you’re just making sense of like the visual representation of the graph. What is like what does this line actually mean? Yeah. So it’s like two to three things going on at once.} -- P01
\end{quote}

\begin{quote}
\textit{I think the whole set of thing was a mess \ldots everybody was trying to ask things at the same time and JooYoung was, you know, like doing his best and still like people were really lost and we have like, sometimes like two or like screen readers at the same time and people. Yeah. That was not ideal.} -- P03
\end{quote}

\begin{quote}
  \textit{I think when I was following with the recording, so second session, I followed it with recording completely. And sometimes I felt like I wish there was kind of instruction given and then give like, I don’t know, like one minute to two minutes time so that people can try it on their own and then kind of catch up to it. So that, so like what I had to do then was like pause the recording to do, do follow the instruction and then see if that works out and then on pause the recording go on. And then I realized that when I joined real time, it wasn’t actually super hard to follow real time. And I think in large parts, I was kind of stop put my jaws on speech on demand and kind of follow it with real display while listening to the jobs that was being shared. So I think I was able to follow through that way.} -- P07
\end{quote}

\subsubsection{Difficulties of Setting Up and Using the Software Environment}
\label{sec:difficulties-setting-up-software-environment}

Another challenge that participants faced was the setup and use of the software environment, which consisted of VSCode as the code editor, R as the programming language, and various R packages for data analysis and visualization. Participants had different levels of proficiency and familiarity with their access technologies (AT), such as screen readers and braille displays, and each AT had its own configuration issues with VSCode and R. Some participants expressed frustration and confusion with the VSCode interface and the R syntax, while others appreciated learning new skills and tools.

\begin{quote}
  \textit{The most challenging was definitely VSCode. Having to deal with the interface. I don’t know. Just think that there were a bunch of gaps between like getting it set up and understanding in the way that it worked. It didn’t click until maybe the third or fourth class that the way our works, the way VSCode works, and the way that we were being taught is it’s not like a functional language in the term that it starts at the top of the bottom. It’s like a scratch pad where you just sort of write where something were any executable code, wherever you wanted in the source file, and then just executed it from there. It was odd to me. And then ultimately, like, it was just, yeah, I just did not like the VScode experience at all.} -- P02
\end{quote}

\begin{quote}
  \textit{The most challenging aspect of the course was to get the environment set up. That was the most challenging for me. That’s my environment was set up. It was easy to follow the course content as well as the instruction.} -- P08
\end{quote}

\begin{quote}
    \textit{I wish I could have done everything ahead though and even though you might not have explained it, maybe going ahead and doing all of the packages, doing everything that we would need so that the instruction could flow. Maybe even if that had been because there was Mac, there were NVDA users and jaws.} -- P09
\end{quote}

\subsubsection{Data Literacy and Confidence}
\label{sec:data-literacy-confidence}

Despite the challenges, participants also reported that the course was a valuable experience for them, as it enhanced their data literacy and confidence. Participants were able to create and interact with various data visualizations, such as bar plots, scatter plots, and box plots, using non-visual techniques, such as sonification and braille representations. Participants also gained a better understanding of the underlying data and the concepts of data analysis and manipulation. Participants expressed a desire for more advanced-level courses on data science and visualization in the future.

\begin{quote}
  \textit{The most rewarding part was the fact that I could not only run and create these various artifacts of data representation myself, but also understand the underlying data. To begin with, there are two parts to what I just said. One is to interact with the software and create these graphs and charts. Then the other part, equally important or maybe more important in some ways, was to understand the underlying data. That had not happened until this. I had interacted with maps and graphs and charts here and there, but then I could never get to the underlying data that was feeding into those charts and graphs.} -- P08
\end{quote}

\begin{quote}
    \textit{I think there is a difference between just manually using the keyboard to produce it versus being asking somebody else to do it. So I think that was really like a meaningful experience.} -- P07
\end{quote} 

\begin{quote}
    \textit{academically, getting first-hand experience with some of the R packages that I wouldn’t otherwise be exposed to. such as BrailleR, R, MAIDR and Tidyverse. \ldots Great. Thank you. That’s a really rewarding.} -- P04
\end{quote}

\begin{quote}
  \textit{The combination of sonification is very cool. Oh, we really appreciate it. \ldots it’s really cool to have a tool like that as a blind person where visually everyone has that tool everyone sighted looks at their outputs \ldots That’s the whole point of visual data analysis. And this does make that visualization come into a position of being almost as efficient for me as it is for sighted people.} -- P05
\end{quote}

%% file: 6-Discussion.tex
\section{Discussion}
\label{sec:discussion}

Teaching data science to blind students in a remote setting, especially delivered by blind instructors, presents unique challenges. This exploration has not only highlighted these challenges but also prompted thoughtful considerations on how to better design accessible learning experiences in the future.

\subsection{Challenges of Real-Time Auditory Processing} 
\label{sec:challenges}

As reported in Section~\ref{sec:challenges-audio-based-learning}, a primary challenge was the difficulty of processing multiple, overlapping sources of auditory information in real time. In a typical coding class, an instructor will likely guide students through code by explaining how particular lines of code work while either typing or highlighting the corresponding code expression. A blind student sitting in a classroom and trying to follow or implement example code in real time has to listen not only to their instructor, but also to the spoken output of their screen reader (often via a single earbud) if they do not have access to a refreshable braille device. They have two competing sources of complex spoken information to attend to. In an online environment where the instructor, too, is using a screen reader to demonstrate code in an accessible way, the problem becomes even more intractable. 

In the classroom environment, a blind student can at least rely on the fact that the two auditory sources are spatially separated to selectively attend to one or the other as the need arises, i.e., to rely on the perceptual phenomenon known as `auditory stream segregation,' whereby distinguishing attributes of auditory sources such as timbre, pitch or spatial location can function to afford each source a unique identity within the auditory environment, an identity that a listener can use to selectively attend to or ignore individual sound sources as when choosing who to listen to at a cocktail party \cite{mcdermottCocktailPartyProblem2009}. But if the instructor, their screen reader, and the student's own screen reader are all arriving through headphones and are not spatially separated or sufficiently differentiated in timbre, the cognitive load can become extremely high. 

A more accessible solution would be to use stereo streaming over headphones, or better to take advantage of the higher spatial fidelity of binaural audio \cite{heerenInfluenceDynamicBinaural2016}. This would allow the instructor's voice, their screen reader and the student's screen reader to be rendered in different locations within a virtual soundscape. Further, as Zekveld and colleagues have noted, source separation by means of timbre is even more effective at reducing cognitive load for a listener than separation by spatial location \cite{zekveldCognitiveProcessingLoad2014}. For the students in our online class, this could be accomplished by employing different synthesizer voices for the screen reader used by the instructor versus those used by the students. Recent advancements in screen readers, such as JAWS and NVDA, facilitate this through their ability to split system sounds and screen reader sounds into separate audio channels, thereby improving the clarity and spatial distinction of auditory information. In our questionnaire, we did not ask participants whether they had chosen to change the voice of their synthesizer to separate it timbrally from that used by the instructor's screen reader. Even if students solved the source segregation problem, there remains a temporal processing challenge as code must be listened to sequentially, without visual scanning abilities. Sufficient gaps must be provided for cognitive processing. As it is impossible to address the individual pacing needs of every student in a real-time classroom environment, we believe that a better approach would be to implement a flipped classroom \cite{reynaEnhancingFlippedClassroom2016} to support self-paced learning.

\subsection{Variability in Assistive Technology Skills}
\label{sec:variability}

Another major barrier was variability in assistive technology skills as noted in Section~\ref{sec:difficulties-setting-up-software-environment}. Although learners were professionals, their experience with screen readers, coding tools, and advanced keyboard shortcuts varied enormously. Pre-teaching exercises or modules focused specifically on onboarding this technical knowledge could more evenly prepare students for advanced data science content. The sheer multitude of interacting mental models is intensely demanding for blind users. Ultimately, successfully teaching data science to blind students in an online environment demands a curriculum intentionally designed for nonvisual learning, as well as attention to the particular needs of a speech-based learning environment \cite{westermannEffectSpatialSeparation2015}. Simply making visual content accessible reactively is insufficient. A proactive, "born accessible" approach considers blind learning needs throughout. Findings suggest that quality learning experiences for blind students require additional scaffolds like flipped classrooms, differentiated streams of information, and extensive onboarding to the complex technical environment \cite{reynaEnhancingFlippedClassroom2016}.

\subsection{Limitations and Future Directions}

Limitations of this initial course are primarily due to its small sample size and brief duration. Furthermore, the course was not structured as a controlled experiment, which means our findings rely on self-reported experiences. To build on this groundwork, future studies should aim for a larger participant base and employ a more rigorous, controlled experimental design to improve the current data science courses. The pilot nature of the course restricted our ability to proactively tackle the two principal challenges highlighted in this study: real-time auditory processing and the wide range in proficiency with assistive technologies. Consequently, future research efforts can be dedicated to the development and empirical evaluation of flipped classroom strategies specifically designed to overcome these barriers. Additionally, future research may warrant a more in-depth exploration of how the data science and visualization education for blind students have unique characteristics compared to the general challenges of non-visual approaches in computer science and programming education identified by the prior literature \cite{stefikBESTPAPERSIGCSE2019,bakerEducationalExperiencesBlind2019}.

%% file: 7-Conclusion.tex
\section{Conclusion}
\label{sec:conclusion}

Overall, the 2-week long online course presented in this paper provides an important first step in exploring the barriers and solutions to making data science and visualization education inclusive for blind students. Participants in the course appreciated the opportunity to learn about data science and visualization and applied their new knowledge to create accessible visualizations (Section~\ref{sec:data-literacy-confidence}). They also provided valuable feedback on the course content and structure, which will be used to improve future iterations of the course. Despite the challenges in remote setting and the short duration, our quantitative and qualitative findings suggest that the course was successful in increasing participants' confidence in their data literacy skills, and in their ability to create accessible visualizations. With a conscious design, data science and visualization can produce new career opportunities rather than obstacles for the blind community. Accessible pedagogy paired with accessible tools can enable their full participation.

%% file: 8-Acknowledgement.tex
\section{Acknowledgement}
\label{sec:ack}

We wish to thank our nine participants who gave so generously of their knowledge and professional experience as data scientists. We also thank the Smith-Kettlewell Eye Research Institute staff for their support in managing the study protocol and for providing administrative support for many aspects of the course.

This work was funded in part by the National Institute on Disability, Independent Living, and Rehabilitation Research (NIDILRR grant number 90 REGE0018) and by the Institute of Information and Communications Technology Planning and Evaluation (IITP) Grant funded by the Korean Government (MSIT), Artificial Intelligence Graduate School Program, Yonsei University, under Grant 2020-0-01361. The MAIDR package used for this course was funded by the Wallace Foundation Grant as part of the first author’s 2022 ISLS Emerging Scholar award, 2023 TeachAccess faculty grant, and the Institute of Museum and Library Services (IMLS) through the Laura Bush 21st Century Librarian Program (grant \#RE-254891-OLS-23).